\documentclass{sigchi}

\CopyrightYear{2018}
\setcopyright{rightsretained}
\doi{https://doi.org/10.1145/3173574.3174014}
\isbn{978-1-4503-5620-6/18/04}
\conferenceinfo{CHI 2018,}{April 21--26, 2018, Montr\'eal, QC, Canada.}
\usepackage{textcomp}

\usepackage{balance}       % to better equalize the last page
\usepackage{graphics}      % for EPS, load graphicx instead 
\usepackage[T1]{fontenc}   % for umlauts and other diaeresis
\usepackage{txfonts}
\usepackage{mathptmx}
\usepackage[pdflang={en-US},pdftex]{hyperref}
\usepackage{color}
\usepackage{booktabs}
\usepackage{textcomp}

\usepackage{microtype}        % Improved Tracking and Kerning
\usepackage{ccicons}          % Cite your images correctly!
\usepackage{todonotes}

\def\plaintitle{Fairness and Accountability Design Needs for Algorithmic Support in High-Stakes Public Sector Decision-Making}
\def\plainauthor{Michael Veale, Max Van Kleek, Reuben Binns}
 
\def\plainkeywords{machine learning, public administration, government, policing, algorithmic accountability.}

\makeatletter
\def\url@leostyle{%
  \@ifundefined{selectfont}{
    \def\UrlFont{\sf}
  }{
    \def\UrlFont{\small\bf\ttfamily}
  }}
\makeatother
\def\pprw{8.5in}
\def\pprh{11in}

\definecolor{linkColor}{RGB}{6,125,233}
\hypersetup{%
  pdftitle={\plaintitle},
  pdfauthor={\plainauthor},
  pdfkeywords={\plainkeywords},
  pdfdisplaydoctitle=true, % For Accessibility,
  bookmarksnumbered,
  pdfstartview={FitH},
  colorlinks,
  citecolor=black,
  filecolor=black,
  linkcolor=black,
  urlcolor=linkColor,
  breaklinks=true,
  hypertexnames=false
}

\usepackage{csquotes}
\newcommand*{\myenquote}[1]{\enquote{{\itshape#1}}}

\newcounter{informantcounter}
\newcommand{\informant}[1]{%
    \ifcsname #1@count\endcsname%
        X\csname #1@count\endcsname%
    \else%
        \stepcounter{informantcounter}%
        \expandafter\xdef\csname #1@count\endcsname{\theinformantcounter}%
        X\theinformantcounter%
    \fi%
}

\begin{document}

\title{\plaintitle}

\numberofauthors{3}
\author{%
  \alignauthor{Michael Veale\\
    \affaddr{University College London}\\
    \affaddr{London, UK}\\
    \email{m.veale@ucl.ac.uk}}\\
  \alignauthor{Max Van Kleek\\
    \affaddr{University of Oxford}\\
    \affaddr{Oxford, UK}\\
    \email{max.van.kleek@cs.ox.ac.uk}}\\
  \alignauthor{Reuben Binns\\
    \affaddr{University of Oxford}\\
    \affaddr{Oxford, UK}\\
    \email{reuben.binns@cs.ox.ac.uk}}\\
}

\maketitle

\begin{abstract}
Calls for heightened consideration of fairness and accountability in algorithmically-informed public decisions---like taxation, justice, and child protection---are now commonplace. How might designers support such human values? We interviewed 27 public sector machine learning practitioners across 5 OECD countries regarding challenges understanding and imbuing public values into their work. The results suggest a disconnect between organisational and institutional realities, constraints and needs, and those addressed by current research into usable, transparent and `discrimination-aware' machine learning---absences likely to undermine practical initiatives unless addressed. We see design opportunities in this disconnect, such as in supporting the tracking of concept drift in secondary data sources, and in building usable transparency tools to identify risks and incorporate domain knowledge, aimed both at managers and at the `street-level bureaucrats' on the frontlines of public service. We conclude by outlining ethical challenges and future directions for collaboration in these high-stakes applications.

\end{abstract}

 \category{H.1.2.}{Models and Principles}{User/Machine Systems} 


\section{Introduction}

Machine learning technologies increasingly form the centrepiece of public sector IT projects, a continuation of existing trends of risk-based regulation~\cite{black2005emergence} given an increasingly `data-driven` flavour. Recently, deployed and envisaged systems have also found themselves under heavy fire from civil society~\cite{nesta2015ml,ainow2016}, researchers~\cite{rsml17,rsdg17,wrr2016bigdata} and policymakers~\cite{goscience2016ai,hocbigdatadilemma,hocrobotics}. These models, often colloquially simply referred to as algorithms, are commonly accused of being inscrutable to the public and even their designers, slipping through processes of democracy and accountability by being misleadingly cloaked in the `neutral' language of technology~\cite{Winner1980}, and replicating problematic biases inherent in the historical datasets used to train them~\cite{barocas2016big}. Journalists, academics and regulators have been recording different examples of algorithmic concerns. Those concerning race have taken centre stage, particularly in the US---ranging from discrimination in recidivism scores used in parole decisions~\cite{propublicamachinebias,Chouldechova_2017} to uneven demographic targeting in systems used for policing on the basis of spatiotemporal crime risk~\cite{selbstpolicing}. 

Bias or opacity are far from newly observed characteristics of computer systems generally~\cite{Friedman:1996,chalmers2003seamful,fischer1991importance}, but these issues have been exacerbated by the rapidly expanding numbers of data-driven information systems entering decision-making processes. They are now increasingly seen within interdisciplinary research communities as highly salient failure modes of these systems, and relevant work seeking to address them is now commonly found in both major machine learning conferences and specialised side-events. The most well-known of the latter, the Workshop on Fairness, Accountability and Transparency in Machine Learning (FAT/ML), has fast become a convening venue and acronym for scholars, initially computer scientists and lawyers but increasingly ethnographers and political scientists, attempting to identify tangible ways by which algorithmic systems can become less undesirably discriminatory and opaque~\cite{Ribeiro:2016,Hajian1:2012,ustunrudin2016,feldman2015certifying,dwork2012fairness}. Techniques in discrimination-aware data mining attempt to define fairness (primarily in reference to anti-discrimination law~\cite{gellert2013law}) and statistically assess it, assure it, or get closer to it~\cite{Pedreshi:2008ej,Hajian1:2012,feldman2015certifying}. Research stemming from the early days of machine learning in expert systems attempts to make algorithmic logics more transparent~\cite{Tickle:1998ed,Wick:1992ko,Montavon2017211,Ribeiro:2016}, or to better understand how users work with inductive systems and build mental models of them~\cite{norman1983some,Kiesler:2002,Tullio:2007}. These tools have attracted attention from governmental bodies, such as the European Commission~\cite{europeancommissionalgoawareness}.

Yet these tools are primarily being built in isolation both from specific users and use contexts. While these are early days for fairness/transparency methods, which are mostly under a decade old~\cite{Pedreshi:2008ej}, the predominant mode of development often involves characterising a problem in a way that might often be at odds with the real world context---such as assuming you have access to the sensitive characteristics you are trying to remove the proxy influence of in a system~\cite{vealebinns}. Yet practitioners are \emph{already} deploying machine learning systems in the public sector, both those designed in-house and those bought from vendors, and are facing immediate, value laden challenges~\cite{coglianese2016regulating}. As observed in studies of clinical decision tools, which often fail to be adopted~\cite{yang2016investigating}, organisational conditions are often missed in design contexts. Indeed fields studying technologies in practice have long struggled to reconcile experimental traditions with the understanding that results from many methodologies fall apart when applied to political, noisy, stressful, complex and contested deployed settings~\cite{galliers1987choosing}. In this paper, we seek to examine existing real-world contexts to assess whether FAT/ML tools might be airdropped from laboratory settings into the messy world of public sector IT with limited adjustment, and if not, in what direction change might be needed. 

To open up questions of practitioner context in relation to fairness and accountability of data-driven systems in the public sector, we interviewed 27 actors working in and with 5 countries' public sectors. These ranged from in-house machine learning modellers to those managing projects, procurement, or contractors delivering and maintaining models. A diverse set of application areas  including taxation, child protection, policing, justice, emergency response and interior security were consciously selected. This was in order to sample widely across practices and challenges emerging in this space rather than try to aim at representative data, which might result in closing-down the issue before it has been sufficiently prised open. The aim was to ascertain how concerns and challenges appear from the perspective of those commissioning and designing these systems on-the-ground. We are only beginning to see this reflective mode in published literature~\cite{marionHART17}, as opposed to attempts to simulate or reverse engineer systems from the outside~\cite{propublicamachinebias,ensignrunaway} (which despite being a highly limited method~\cite{seaver_reverse}, is one of few remaining choices where systems are resolutely closed).

Following further background and our methodology, we highlight and elaborate on core themes from undertaken interviews. Themes are grouped into those relating to internal users and stakeholders of the decision support systems and those relating to external users, stakeholders and decision subjects. We then discuss some central interdisciplinary challenges that emerge from these findings that we believe to be of importance to both design communities and to broader sets of scholars, policy-makers and other practitioners.

\section{Background and motivation}

Significant prior research interest exists concerning high-stakes, operational decision support. Decision-making in clinical contexts was an early focus in HCI and human factors due to its intrinsic importance to lives and livelihoods, the possibility of codification in highly specialised areas, and the development (and funding) of a range of relevant, high-profile expert systems. Canonical domains in human factors have also focussed on even more dramatic life-and-death situations, such as issues around decision support in airplane cockpits~\cite{mosier98} and air traffic control~\cite{hopkin1995human}.

While these studies are important to reflect and build upon, much algorithmically-informed public sector decision-making is not best conceived as simply a direct continuation of these specific challenges. The reason for this stems from an assumption of most work on decision-support systems: that it is relatively straight-forward to measure whether a design intervention would `reliably improve human decision-making, particularly in real-world contexts'~\cite{zhang2015designing}. Does the patient improve, or at least receive the same treatment that would have been recommended by a top specialist? Does the plane glide or plummet? These situations, many of which can be characterised as `safety', represent problems structured so that there is high consensus on both the \emph{ends} (goals such as remaining in flight) and the \emph{means} (well-understood steps that achieve those goals).

Those managing public policy problems rarely have the luxury of the settled consensus on ends and means some engineers are used to~\cite{hoppe2011governance}. If a police department turns to a machine learned predictive model to anticipate crime risk in different parts of a city, they face a range of debates. A desired end might be to treat all crime equally. But does that imply police should focus resources on areas of high crime at the expense of those with low crime, to maximise total arrests? Or does it mean that a crime in a low-risk area is just as likely to be intervened in as a crime in a high risk area? Areas conceived of as `high risk' are rarely distributed at random, coupled instead to communities with different demographic or vulnerability distributions. The means are also unclear. Should models be used to increase preventative measures, such as community policing, or to heighten response capacity after crimes have been reported? Of course, decisions are rarely this binary, but that does not mean they are settled. At first glance, a problem might seem like a clinical decision. Scratching the surface, myriad subjective choices quickly arise.

Compounding this, public sector IT projects are heavily resource constrained, path-dependent given existing infrastructures, and prone to failure from poor initial scoping~\cite{DunleavyDEG}. Their information systems are notorious for crossing scales and chains of accountability, engendering many practical challenges relating to performance and maintenance~\cite{le2010across}. Uptake of models in government is as much a social process as it is a technical one \cite{kolkman2016build}. Furthermore, many of the important dynamics are somewhat hidden from public view. As noted in the public administration literature, values in public service are primarily exercised in the discretionary spaces between the rules rather than in the high level political rule-making processes themselves~\cite{Jorna:2007wl,Lipsky2010}. Many ethical issues are only likely to surface far down the road of creation, procurement or deployment of a system, after many choices are baked-in~\cite{forsythe1995using}. This chimes with a common observation from clinical decision support---that designed tools, while promising in laboratories, often fail in practice in a variety of ways~\cite{Kawamoto765}. Only recently have HCI researchers in these settings begun to investigate \textit{institutional} contexts behind these failures~\cite{yang2016investigating}, noting primarily that airdropped decision support systems often fail to embrace the richness of the clinical context. 

A parallel concern motivates this work: that nascent FAT/ML tools, such as debiasing or transparency systems, will also fail unless contextual challenges are understood early on. We are not aware of other research  considering how value-laden concerns manifest and are coped with in real, public sector decision support settings---something that speaks more to the recent increase in these types of tools than to the novelty of our broad approach. Indeed, while there has been a surge in researchers considering the social implications of `algorithms', we are wary that issues are being framed from afar, rather than in collaboration with those facing them and perhaps understanding aspects that may not always be immediately apparent. As a result, we do not seek to populate or validate any of the young and primarily unverified theoretical frameworks, but lay empirical foundations for a more grounded approach that might enable more positively impactful work in the future. In particular, in summarising findings for this work we sought to highlight aspects infrequently discussed in or omitted entirely from relevant contemporary research discussions. The section that follows explains how we sought to do that.

\section{Method}

Twenty-seven individuals agreed to be interviewed in late 2016---predominantly public servants and attached contractors either in modelling or project management. Each individual was interviewed only once, either in person (seventeen interviews) or on the telephone (ten interviews). They were all undertaken with one interviewer, and each lasted between forty and sixty minutes. Just over one-fifth of informants were female. Interviewees worked in one of five OECD countries located over three continents. It was decided in the initial ethical approval for this work not to publicly name the countries in order to reduce the risk of informant identification, but we will note that the US was not one of the countries included.

Informants were identified with an \emph{ad hoc} sampling approach. This was chosen for several reasons. Firstly, at this relatively early stage of deployment, projects are emerging without central mandates---no coordinating body was identified to have a reliable compiled register of activities. Indeed central agencies occasionally shared registers that turned out to be a poor representation of on-the-ground  activities. Secondly, we sought as many perspectives as possible from within public sector organisations deploying machine learning for decision-support today, and felt this was best achieved by looking across sectors to very different types of agencies and bodies. To recruit participants, projects and contacts were assembled from grey literature, freedom of information requests (both actively made and through platforms such as \emph{WhatDoTheyKnow}), snowball sampling, direct inquiries with organisation contacts, and the use of news databases including \textit{Factiva} and \textit{LexisNexis}. Terms including \emph{predictive modelling}, \emph{entity-level prediction}, \emph{predictive analytics} and \emph{machine learning} were entered into these databases and public document repositories. Participants additionally played an important role in sampling themselves, and were usually willing and often even eager to flag colleagues in other domestic or foreign agencies working on projects they felt would benefit the study. Similarly to challenges arranging interviews with societal `elites', candidacy for interviews 'often cannot be planned for adequately in advance of the project; rather, it emerges as part of the fieldwork'~\cite{odendahl2002interviewing}.

Because of the open-ended nature of the sampling, the varied nature of the roles (particularly across sectors), and the many different systems concerned, it was neither possible nor helpful to stick to a rigid script. Instead, the approach taken was similar to other open-ended work in policy research, involving prompting the participant to not only outline their role but explain the process behind the development and maintenance of the project.\footnote{See~\cite{page2005policy}, who conducted 128 interviews in the UK civil service to understand the nature of policy work, asking only  `what do you do?' and `how do you come to be in this job?'.} First, the purpose of the study was explained to participants, at which point any ambiguities could be resolved. Following that, participants were asked about their role (and history of roles) in this area, then to give a high level outline of relevant project(s) and a more detailed view on their position within them. They were then steered at opportune moments in the discussion towards topics of fairness and accountability, effectiveness and complexity/robustness (mirroring the public sector values framework introduced by~\cite{Hood:1991vk}). At times, this steering was unnecessary and avoided, particularly as the nature of the study was made clear to participants: many already had considered these issues during their job, albeit often under different names. The other main prompt used to elicit relevant insights, particularly where participants had not considered their job in the above framing before, was to ask whether `anything surprising or unexpected' had happened to them in relation to their work, such as a deployed model. This was especially useful in eliciting institutional events, or novel incidences of model failure.

Conversations were not taped. While that might have been desirable, recording audio of individuals discussing sensitive public sector work is extremely difficult. Bodies commonly disallow it when individuals are not spokespersons for the organisation, precluding it as an interview approach, more so where new technologies are involved, and questions asked are likely to be totally new. Even where taping is permitted, it can risk inhibiting openness and frankness in discussions. These politically-charged contexts pose methodological restrictions infrequently seen in HCI, but frequently encountered by public administration researchers, and we follow methodological practices developed in this field~\cite{page2005policy}. These are further exacerbated here by fear of negative media coverage---both journalists and academics in this field have exhibited a recent taste for algorithmic `shock stories'. Instead, verbose notes were continuously taken with the aim of authentically capturing both interviewees' tone, phrasing and terminology, as well as the core points they explained. Where longer continuous answers were given, interviewees kindly paused for note-taking purposes. Notes were typed up by the interviewer, always on the same day as the interview took place and often immediately following. Some highly context-specific terminology, such as geographic subunits or revealing was substituted with equivalent generic alternatives to increase the difficulty of project re-identification. Handwritten notes were then destroyed in line with data protection and the study's ethical approval.

To analyse the interviews, open coding was used (\textit{NVivo 11 for Mac}), with codes iteratively generated and grouped concerning the challenges and coping mechanisms observed. These were then iteratively grouped according to a public sector values framework from the public administration literature ~\cite{jorgensen2007public} as a thematic organisational principle.

\section{Findings}

In this section, we summarise some of the key themes from the interviews undertaken. They are split into two broad sections: those concerning internal actors and their relation to the algorithmic systems, such as other departments, and those concerning external actors, such as decision subjects.

\subsection{Internal actors and machine learning--driven decisions}

The first category of themes relate to discussions of how the deployed systems are connected to and perceived by a range of internal actors. Many of the issues around algorithmic transparency so far have focussed on external algorithmic accountability and transparency-based rights, such as a `right to an explanation'~\cite{edwardsveale}, although broad reasons to make systems transparent and interpretable exist~\cite{Lipton:2016tw}. Yet within organisations there are a wide array of reasons for understanding data and models.

\subsubsection{Getting individual and organisational buy-in}

Informants reported a need to use different approaches to clarify the workings of or process behind machine learning powered decision-support systems for internal actors. Some of these were strategic actors in management positions, either the clients of external contractors or customers of internal modelling teams.

Several interviewed practitioners noted that this organisational pressure led them to make more `transparent' machine learning systems. Detection systems for fraudulent tax returns illustrated this. The analytics lead at one tax agency [\informant{taxlead1}] noted that they \myenquote{have better buy-in} when they provide the logic of their machine learning systems to internal customers, while their counterpart in another tax agency [\informant{taxlead2}] described a need to \myenquote{explain what was done to the business user}. Both these individuals and modellers around them emphasised they had in-house capability for more complex machine learning systems, such as support vector machines or neural networks, but often chose against them for these reasons. Instead, many of the systems that ended up being deployed were logistic regression or random forest based.

Some saw transparency in relation to input variables more than model family. One contractor that constructed a random-forest based risk score for gang members around knife crime on behalf of a global city's police department [\informant{police-c1}] described an \myenquote{Occam's razor} process, where they started with 18,000 variables, working down to 200, then 20, then 8---\myenquote{because it's important to see how it works, we believe}. To steer this, they established a target percentage of accuracy with the police department \textit{before} modelling---around 75\%---which they argued helped them avoid trading off transparency. When users of analytics are not \myenquote{confident they know what a model is doing}, they \myenquote{get wary of picking up protected characteristics}, noted the modelling lead at tax agency [\informant{taxmodel1}]. To make this more transparent, the police contractor above [\informant{police-c1}] would \myenquote{make a model with and without the sensitive variables and see what lift you get in comparison}, presenting those options to the client to decide what was appropriate.

Another issue raised by several modellers was the difficulty in communicating the performance of designed systems. One modeller in a regional police department [\informant{police-a2}] was designing a collaborative system with neighbouring police departments to anticipate the location of car accidents. They noted that

\begin{quote} \itshape
 We have a huge accuracy in our collision risk, but that's also because we have 40 million records and thankfully very few of them crash, so it looks like we have 100\% accuracy---which to the senior managers looks great, but really we only have 20\% precision. The only kind of communication I think people really want or get is if you say there is a 1/5 chance of an accident here tomorrow---that, they understand.
\end{quote}

An analytics lead at a tax department [\informant{taxlead2}] faced parallel issues. When discussing the effectiveness of a model with clients, he would often find that \myenquote{people tend to lose faith if their personally preferred risk indicators aren't in a model, even without looking at performance of results.} 

Performance was often judged by the commissioning departments or users based on the additional insight it was thought to provide, compared to what they thought to be known or easily knowable. There was a tension between those who were seeking insight beyond existing processes, and those seeking efficiency/partial automation of current processes. One contracted modeller for a police department [\informant{police-c1}] noted that during modelling, they \myenquote{focussed on additionality. The core challenge from [the police department] was to ascertain whether the information we could provide would tell them things they did not already know. How would it complement the current way of doing things?} Yet another case, an in-house police modeller [\informant{police-b3}] noted that a focus on additionality by the users of the system often clouded the intended purpose of the software in the first place.

\begin{quote} \itshape
What we noticed is that the maps were often disappointing to those involved. They often looked at them and thought they looked similar to the maps that they were drawing up before with analysts. However, that's also not quite the point---the maps we were making were automatic, so we were saving several days of time.
\end{quote}
 
\subsubsection{Over-reliance, under-reliance and discretion}
% other research that i might want to cite... people ignoring encow? 
Over and under-reliance on decision support, extensively highlighted in the literature on \textit{automation bias}~\cite{Skitka:1999il,Dzindolet:2003bl}, featured considerably in informants' responses. A lead machine learning modeller in a national justice ministry [\informant{justice1}], whose work allocates resources such as courses within prisons, described how linking systems with professional judgement \myenquote{can also mean that [the model output is] only used when it aligns with the intuition of the user of the system}. To avoid this, some informants considered more explicitly how to bring discretion into decision-support design. A lead of a geospatial predictive policing project in a world city [\informant{police1}] noted that they designed a user interface

\begin{quote} \itshape
to actively hedge against [officers resenting being told what to do by models] by letting them look at the predictions and use their own intuition. They might see the top 3 and think `I think the third is the most likely' and that's okay, that's good. We want to give them options and empower them to review them, the uptake will hopefully then be better than when us propellorheads and academics tell them what to do...
\end{quote}

Model outputs were not treated similarly as decision support in all areas. The former lead of a national predictive policing strategy [\informant{police-b2}] explained how they saw discretion vary by domain.

\begin{quote} \itshape
We [use machine learning to] give guidance to helicopter pilots, best position them to to optimise revenue---which means they need to follow directions. They lose a lot of flexibility, which made them reluctant to use this system, as they're used to deciding themselves whether to go left or right, not to be told `go left'! But it's different every time. There were cases where agents were happy to follow directions. Our police on motorcycles provide an example of this. They were presented with sequential high risk areas where criminals should be and would go and apprehend one after another---and said ``yes, this is why we joined, this is what we like to be doing!'' The helicopters on the other hand did not like this as much.	
\end{quote}

Also faced with a list of sequential high risk activities, this time relating to vulnerability of victims, the analytics lead at one regional police department [\informant{police-a1}], sought advice from their internal ethics committee on how to use the prioritised lists their model outputted.

\begin{quote} \itshape
We had guidance from the ethics committee on [how to ethically use rank-ordered lists to inform decision-making]. We were to work down the list, allocating resources in that order, and that's the way they told us would be the most ethical way to use them... It's also important to make clear that the professional judgement always overrides the system. It is just another tool that they can use to help them come to decisions.
\end{quote}

\subsubsection{Augmenting models with additional knowledge}

Informants often recognised the limitations of modelling, and were concerned with improving the decisions that were being made with external or qualitative information.  A lead of a national geospatial predictive policing project [\informant{police-b1}] discussed transparency in more social terms, surrounding how the intelligence officers, who used to spend their time making patrol maps, now spent their time augmenting them.

\begin{quote} \itshape
We ask local intelligence officers, the people who read all the local news, reports made and other sources of information, to look at the regions of the [predictive project name] maps which have high predictions of crimes.  They might say they know something about the offender for a string of burglaries, or that building is no longer at such high risk of burglary because they local government just arranged all the locks to be changed. [...] We also have weekly meeting with all the officers, leadership, management, patrol and so on, with the intelligence officers at the core. There, he or she presents what they think is going on, and what should be done about it. 
\end{quote}

Other types of knowledge that modellers wished to integrate were not always fully external to the data being used. In particular, information needs also arose linked to the primary collectors of training data. One in-house modeller in a regional police department [\informant{police-a2}], building several machine learning models including one to predict human trafficking hotspots, described how without better communication of the ways the models deployed worked, they risked large failure.

\begin{quote} \itshape
Thankfully we barely have any reports of human trafficking. But someone at intel got a tip-off and looked into cases at car washes, because we hadn't really investigated those much.\footnote{Modern slavery is a problem in the car wash industry~\cite{carwash}.} But now when we try to model human trafficking we only see human trafficking being predicted at car washes, which suddenly seem very high risk. So because of increased intel we've essentially produced models that tell us where car washes are. This kind of loop is hard to explain to those higher up.
\end{quote}

Similarly, external factors such as legal changes can present challenges to robust modelling. A modeller in a justice ministry building recidivism prediction systems noted that while changes in the justice system were slow, they were still \myenquote{susceptible to changes in sentencing, which create influxes of different sorts of people into the prison systems.} These kinds of rule change are unavoidable in a democratic society, but awareness of them and adequate communication and preparation for them is far from straightforward.

\subsubsection{Gaming by decision-support users}

`Gaming' or manipulation of data-driven systems, and the concern of this occurring if greater transparency is introduced, is  often raised  as an issue in relation to the targets of algorithmic decisions. This will be discussed in a following section. Yet types of \textit{internal gaming} within organisations have received considerably less treatment by those concerned about value-laden challenges around algorithmically informed decisions. This is despite how internal gaming is extensively highlighted in the public administration literature in relation to targets and the rise of New Public Management~\cite{Bevan:2006vk}, a broad movement towards `rationalisation' in the public sector that clearly affected informants around the world. 

One tax analytics lead [\informant{taxlead2}] worried that releasing the input variables and their weightings in a model could make their own auditors investigate according to their perception of the model structure, rather than the actual model outputs---where they believed that bias, through fairness analysis, could ostensibly be controlled.

\begin{quote} \itshape
To explain these models we talk about the target parameter and the population, rather than the explanation of individuals. The target parameter is what we are trying to find---the development of debts, bankruptcy in six months. The target population is what we are looking for: for example, businesses with minor problems. We only give the auditors [these], not an individual risk profile or risk indicators [...] in case they investigate according to them.
\end{quote}

Additionally, some tax auditors are tasked with using the decision-support from machine learning systems to inform their fraud investigations. Yet at the same time, the fraud they discover feeds future modelling; they are both decision arbiter and data collector. The effect these conflicting incentives might have on a model were highlighted by a different tax agency [\informant{taxlead2}], as when auditors accumulate their own wages, \myenquote{[i]f I found an initial [case of fraud], I might want to wait for found individuals to accumulate it, which would create perverse incentives for action}.

\subsection{External actors and machine learning--driven decisions}

The second theme focusses on when informants reflected upon value concerns that related to both institutional actors that were outside their immediate projects, or that were at a distance, such as subjects of algorithmically informed decisions.

\subsubsection{Sharing models and pushing practices}
Scaling-up is an important part of experimentation. This is particularly the case in public sector organisations replicated by region---while some of them, particularly those in the richest or densest areas, can afford to try new, risky ideas with the hope of significant performance or efficiency payoffs to outweigh their investment, for smaller or poorer organisations that economic logic does not balance. The latter set of organisations are more reliant on the import and adaptation of ideas and practices from more well-resourced sister organisations (which could also be abroad) or from firms. Yet in practice, this is challenging, as machine learning systems also come imbued with very context specific assumptions, both in terms of the problem they are attempting to model, and the expertise that surrounds the decision-making process each day it is used. A modeller and software developer in  a spatiotemporal predictive policing project [\informant{police-b3}] emphasised the challenges in scaling up these social practices, as they were not as mobile as the software itself.

\begin{quote} \itshape
If you want to roll out to more precincts, they have to actually invest in the working process to transform the models into police patrols. To get more complete deployment advice... it takes a lot of  effort to get people to do that. What you see is that other precincts usually---well, sometimes---set up some process but sometimes it is too pragmatic. What I mean by this is that the role of those looking at the maps before passing them to the planner might be fulfilled by someone not quite qualified enough to do that.
\end{quote}

Similar sentiments were also echoed by individuals in national tax offices, particularly around the `trading' of models by large vendors. One tax analytics lead [\informant{taxlead2}] in a European country expressed concerns that another less resourced European country was being sold models pre-trained in other jurisdictions by a large predictive analytics supplier, and that they would not only transpose badly onto unique national problems, but that the country interested in purchasing this model seemed unprepared to invest in the in-house modelling capacity needed to understand the model or to change or augment it for appropriate use.

\subsubsection{Accountability to decision subjects}

Interpretable models were seen as useful in relation to citizens. One lead tax analyst [\informant{taxlead2}] described how transparency provided \myenquote{value-add, particularly where an administrative decision needs explaining to a customer, or goes to tribunal}. They noted that \myenquote{sometimes [they] justif[ied] things by saying here are the inputs, here are the outputs} but they were \myenquote{not really happy with that as an ongoing strategy.} Yet on occasion, more detailed explanations were needed. The same informant recalled an incident where a new model, in line with the law, was flagging tax deductions to refuse that were often erroneously allowed to some individuals in previous years. Naturally, many people called in to complain that their returns were not processed as expected---so the tax agency had to build a tool to provide call centre operators with client-specific explanations.\footnote{It was unclear if this system was machine learning or rule-based.}

Other organisations focussed on providing knowledge of the system to other interested parties, such as media organsiations. One national predictive policing lead [\informant{police-b1}]  explained how they found it difficult to have discussions around equity and accountability with police officers themselves, who are often narrowly focussed on \myenquote{where they think they can catch someone}, and have less capacity or incentive to devote time and energy to frame broader questions. Instead, this police force would invite journalists over twice a year to see what the predictive teams \myenquote{do, how [the algorithms] work, and what we are doing}. Several public sector organisations using machine learning systems already publish information about the weights within their model, the variable importance scores, or record ethical experiences and challenges in the modelling process~\cite{tollenaar2016,mooreOASys,marionHART17}.

\subsubsection{Discriminating between decision-subjects}

Discrimination has taken centre-stage as the algorithmic issue that perhaps most concerns the media and the public. Direct use of illegal-to-use protected characteristics was unsurprisingly not found, and interviewees were broadly wary of directly using protected characteristics in their models. Input data was seen as a key, if not the only point of control, but the reasons and the logics behind this varied. A lead of analytics at a national tax agency  [\informant{taxlead2}] noted that \myenquote{if someone wanted to use gender, or age, or ethnicity or sexual preference into a model, [they] would not allow that---it's grounded in constitutional law.} In one case guidance was to be released clarifying forbidden variables, but made no difference as the tax agency was already compliant [\informant{taxlead1}]. Even when characteristics were found to be legally permitted after consultation with lawyers (characteristics are not protected in all contexts), they might still have been avoided. Informant [\informant{taxmodel1}], a lead modeller in a tax agency, noted that they have an informal list \myenquote{of variables that [they] don't feed into models}, which included age and location, both of which were legally permissible in their context. Location was avoided by this informant because even though different cities have different tax fraud risks, they \myenquote{don't usually want to investigate on those grounds.} In other cases, home location was avoided as it was a \myenquote{proxy for social deprivation}, in the words of the lead modelling a justice ministry [\informant{justice1}] .

Occasionally, there would be pressure to use protected characteristics to increase predictive power. The same justice modeller [\informant{justice1}], noted that \myenquote{we had feedback from a senior [foreign nationality, omitted] academic in this space on our [criminal justice] model, noting that `if you've got something as predictive as race is, why aren't you using it?' Many of [this experts' deployed] models do, but it's an ethical decision in my mind and this is the route we've taken.} Relatedly, they were also concerned by how the proxy outcome that could be measured (conviction) related to sensitive variables, rather than the outcome variable of real interest (offending).

\begin{quote} \itshape
Race is very predictive of re-offending, [but] we don't include race in our predictive models [...] we are aware that we are using conviction as the proxy variable for offending, and if you do this then you can get into cycles looking at certain races which might have a higher chance of being convicted, and train models on this data instead. That would mean you're building systems and catching people not based on the outcome, but on proxy outcomes.
\end{quote}

\subsubsection{Gaming by decision-subjects}

It is commonly expressed that extensive transparency of algorithms to the public might encourage system gaming~\cite{natureed2016}, and this is brought to bear as a justification for opacity. Correspondingly, external gaming was raised as an issue by some informants. One contractor developing predictive policing software for a world city [\informant{police-c1}] noted that concerns in his sector concerned \myenquote{criminal gangs that might send nine guinea pigs through the application process looking for loopholes to get arrested, just to find a tenth that highlights a way they can reliably get passports from under the noses of the authorities.}. An analyst from a large NGO working in collaboration with the police on developing a predictive system to detect child abuse [\informant{ngo1}] noted that \myenquote{it's much harder to game when you've linked up lots of different aspects, education and the like.}, although their colleague [\informant{ngo2}] warned that they were concerned about many of the usual sophisticated practices being used to game ML-supported systems, such as \myenquote{turning professionals against each other} or the \myenquote{strategic withholding of consent at opportune moments}. The analytics lead at one tax agency [\informant{taxlead1}] explained that while they would publicly share the areas they were interested in modelling tax fraud for, such as sectors or size, they were \myenquote{primarily concerned that if the model weights were public, their usefulness might diminish}.

Other incidents resembled gaming---and could feasibly be interpreted as such---but served more to demonstrate the current fragility of models towards concerted attempts to change them. A modeller at a police department [\informant{police-a2}] noted, in relation to a model they had built to pre-empt when the force should ensure they had the most staff available to deal with missing persons, that

\begin{quote} \itshape
There's one woman who calls in whenever her kid is out after 10pm. She then calls back about 30 minutes or so later to say that everything is fine, or we follow up with her. But then it looks like in the model that kids always go missing at 10pm, which obviously is a bit misleading. In the end I had to manually remove her from the model to remove the spurious pattern.
\end{quote}

While in this case, the model failed---resembling an \textit{availability attack}, to draw on the adversarial machine learning literature---this might not always be the case. Indeed, models might not fail in obvious ways, or might even be subject to attacks designed to change them in targeted ways~\cite{huang2011adversarial}. Even where an attack is not planned, simply responding to decisions informed by the model---such as patrol patterns---might look like gaming, or at least a game of cat-and-mouse. The police lead on a geospatial predictive policing project for a world city [\informant{police1}] noted this in their own system. While it wasn't clear whether they were just removing the lowest hanging fruit or criminals were responding, in response, they linked a further feedback effect to try to compensate for the performance loss.

\begin{quote} \itshape
The highest probability assessments are on the mark, but actual deployment causes displacement, dispersion and diffusion, and that throws the algorithm into a loop. You have to remodel, though typical patterns of unresponded-to crime are predicted well [...] we decided to re-evaluate learning every 2--3 weeks, pull in all sorts of other variables, such as feeding it with what police were deployed, what they did---I've never seen this in other similar systems. In the first four weeks of trialling it out, the probability of being correct just tanked [...] in the 3rd update, it started to figure shit out.
\end{quote}

\section{Issues ahead}

In this section we draw upon the responses from informants to point to several `grand challenges' for high stakes, algorithmically-informed public decisions. This is not a comprehensive list or typology. Instead we are seeking to emphasise areas where we believe vigorous discussion in the FAT/ML, HCI, human factors, critical data studies, information systems and public administration communities, among others, is lacking and needed. While these are not intended as direct implications for design, we see opportunities for design within each of them, as well as opportunities for many other fields and types of knowledge to be brought to bear.

\subsection{`The probability of being correct tanked': Data changes}

Data in the public sector is usually collected, categorised and cleaned for primarily operational reasons, such as recording who has been arrested, calculating tax bills, or delivering mail---not for modelling. While increasing emphasis is now put on secondary uses of data~\cite{admindatanetwork2012}, primary needs remain primary. Upstream changes in the logic of collection---as was the case above when investigative patterns led to a human trafficking risk model becoming a car-wash detector~\cite{carwash}---can have significant downstream effect. Particularly where models are quietly being developed or piloted, or are located in a different part of the organisation from data collection efforts, it is easy for changes in practices to occur without those responsible for model performance to be aware of them. Accountability becomes difficult to trace in these situations. As Nick Seaver puts it, these systems are `not standalone 
little boxes, but massive, networked ones with hundreds of hands reaching into them'~\cite{seaver2013knowing}. Accountable systems should to be internally accountable, else it would appear to be difficult for external accountability to either make sense or be sustained.

Data can also change \emph{because} of the model rather than only in spite of it. Where models allocate the same resources that collect data then they are directly influencing the future sampling of their training data~\cite{sculley2015hidden}. Sending police officers to areas of high predicted crime is an example of this. In the worst cases, the model can have a polarising effect: directing police resources disproportionately to areas with slightly higher crime risk will, without corrections, skew future training data collection in those area, which might be demographically or socioeconomically disproportionate~\cite{ensignrunaway}. In other cases, effects might be more subtle but of equal importance, and might cause particular failures to occur in unforeseen ways. If individuals react to try and influence or game a system---as the example stories above indicate is certainly possible---then the future population distribution becomes a function of past model decisions or structure. Little work has focussed on this so far, particularly on the impacts of these on the research into statistical fairness and non-discrimination properties, which broadly implicitly assume stationarity in their problem set-up. This is also a topic not substantively covered in existing literature, which is largely founded on data collected online, such as in the process of optimising advertising revenue. The adverts you are delivered might slowly change your behaviour, but each one can hardly be thought to have a significant impact. This is not the case in the public sector. As [\informant{police1}] recalled above when discussing crime dispersion, feedback effects in practice can be so strong that they make models rapidly fail. The effect of this property on fairness and accountability in systems has yet to be properly unpacked and explored in context.

How to respond to concerns around shifting data? To some extent, the problem is highly interpersonal. The notion of a \emph{visibility debt} has received some attention from engineers of both machine learning and traditional software engineering~\cite{visibilitydebt,sculley2015hidden}. To the upstream data collectors, there are \emph{undeclared users} of their data streams. To the downstream users, there are individuals exerting influence over their system that that might not even be aware that such a system exists, particularly when data is collected in a decentralised manner by, say, auditors or police patrol officers. This problem is only going to be exacerbated as more models are made using the same upstream data sources, and bilateral communication becomes more and more challenging. Better communication might help, but must overcome difficult hurdles of explaining to upstream actors the kind of changes that matter downstream, and the kind that don't, in ways that they not only understand (as they might be relatively statistical) but that they can identify and act on within their roles. This is all compounded by how changing upstream data collection might not be an explicit act at all, but one emerging from cultural change or use of discretion. This is emphasised in the importance of so-called `street-level ministers' in the public administration literature, which points out how formal rules are only part of the picture, and that many day-to-day choices in the grey zones are made by bureaucrats at the frontlines of public service~\cite{Lipsky2010}. Where change does occur, managers might not notice it, as in their day-to-day roles or through their monitoring and evaluation tools, they only see part of the picture~\cite{buffat2015street}.

A second approach would assume communication failure is inevitable, pushing instead a focus on the changing data itself. This would involve concept drift detection, sets of techniques designed to automatically detect shifts in distributions potentially relevant to a modelling task. Concept drift detection, particularly in complex real-world contexts, is difficult and daunting theoretically, let alone practically~\cite{QuinoneroCandela:2009ud,Gama:2014ch}. Some of the more recent reviews in the field call for the integration of domain knowledge in order to discern relevant drift~\cite{Gama:2014ch}, yet there are few, if any well-explored methods for doing this.

\subsection{`Always a person involved': Augmenting outputs}

While we hear horror stories of the results of algorithms unquestioningly replacing swathes of existing analytical practice and institutional knowledge, our informants' experiences do not reflect that. Many organisations interviewed here have well-developed routines for augmenting algorithmic outputs, such as crime maps, with contextual data using manual analysts. As one informant described, their `predictive policing' system was not supposed to bring in shocking new insights, but relieve analysts from the slog of generating maps so that they could get on with more advanced work. How algorithmic systems are examined day-to-day and how humans enter `the loop' of decision-making at different stages is an important area for future design focus. There are many points for intervention in a decision support system outside of the modelling process---for example, in the training data (many systems attempting to make fairer machine learning system intervene at this point~\cite{kamiran2012data,feldman2015certifying}) or after the model has been generated~\cite{kamiran2010discrimination}, such as the stage between model output and map dissemination~\cite{chen2005visualization}. Particularly in this latter stage, design interventions are likely to be key. If a statistical definition of fairness is reached, it may be possible to make a `fair' model, for example by introducing fairness constraints to optimisation. This provides no guarantees about decision-support being interpreted fairly. Designers should not just consider how to design artifacts such as maps to promote fairness, but should also do so in contexts imagining that models have been `scrubbed' of certain types of bias, to understand if this introduces any additional effects. In the messy outside world, these efforts may interact, and it is not guaranteed that the sum of two good efforts is also effective.

Taking this areas forward will likely require building upon and rethinking traditional knowledge elicitation techniques. Effective knowledge elicitation, as part of the hot topic of knowledge acquisition in heady days of expert systems, was thought to be a foundational building block of AI~\cite{COOKE1994801,hoffman2008human}. With the inductive, data-driven turn, we may need to rediscover it as something which constrains and augments patterns learned from data, less around tough or rare cases~\cite{hoffman1998use} as much as around contentious, value-laden ones. This will require very different sorts of prioritisation and elicitation methods than developed so far, and seems a promising and urgent avenue for future research.

\subsection{`When it aligns with intuition': Understanding discretion}

It is commonly claimed that people over-rely on algorithmic systems, or increasingly consider them neutral or authoritative~\cite{boydforum}. We do not claim this is not an issue---but according to the informants in this project, this framing is one-dimensional. In particular, if and how individuals trust and rely on decision-support systems seems highly contextual in nature. The design strategies used to improve uptake of these systems, such as presenting prioritised lists or options, are understudied in relation to how these affect the mental models constructed by those using these systems day-to-day.

The way that different tasks, stakes or contexts mediate these effects is even less studied.  We might expect there to be a difference in the perception of `neutrality' of algorithms between those that direct police helicopters and those that flag children at risk of abuse; two very different tasks. We might not expect however, as informants reported, there to be a significant difference in the way algorithms were considered by helicopter pilots versus by police motorcyclists. Research in risk perception by helicopter pilots has found additional disparities between experienced and inexperienced users which is also worth unpacking in this context~\cite{thomson2004aviation}. Ultimately, to make blanket and somewhat alarmist statements about how algorithms are or are not being questioned is likely to alienate practitioners who recognise a much more nuanced picture on the ground, and hinder co-operation in this space between researchers and those who would benefit from research uptake.

As well as the demographics and contexts of when algorithms are trusted more or less on aggregate, we might be interested in patterns of over- or under-reliance \emph{within} individuals or use settings. If users of decision-support choose to ignore or to follow advice at random, we may not be wholly concerned with this, or at least our concern might centre on the dimension of increasing adherence. Yet if there are \emph{systematic} biases in the way that advice is or is not used---particularly if they result in individuals holding different protected characteristics being treated differently---then this may create cause for alarm, or at least merit further study. Assuming a system can be `scrubbed' of bias and then forced onto users to obey is clearly not what will happen in real world deployments. 

Lastly, researchers considering human factors in computer security have emphasised `shadow security practices', which consist of `workarounds employees devise to ensure primary business goals are achieved' and `reflect the working compromise staff find between security and ``getting the job done'''~\cite{Kirlappos:2015}. Similarly, studies of fairness and accountability in socio-technical systems must incorporate an assumption that there will be a mixture of technological resistance and ad-hoc efforts, which, similarly to the findings in human factors of security, will surely be `sometimes not as secure as employees think.' You can't engineer ethics, and you can't expect some individuals not to try, rigorously or not, to uphold it in ways they see fit. It is a useful heuristic to assume systems are trained on `pure' streams of data and then must be cleaned of bias downstream, but in real data collection environments, even upstream actors in the data collection process attempt to work in the discretionary places computer systems allow (and create) to inject fairness where they see fit~\cite{Jorna:2007wl}.

\subsection{`I'm called the single point of failure': Moving practices}

Most of the work in discrimination-aware data mining involves statistical assurance of fairer systems, or the installation of interfaces to make them more transparent. Most of the experiences of informants in this study were the opposite---social detection of challenges and social solutions to those challenges, none of which were mathematically demonstrated to work, but which organisationally at least were perceived to be somehow effective. Managing these challenges will require a balance between the two that has seldom been effectively struck. It seems unlikely that statistical practices could exist without the social practices, or the other way around.

This means that how the social practices are developed, maintained and transferred across contexts or over time is important to consider. Public sector bodies are under the constant shadow of their core quantitatively trained staff being poached, moving agencies, or leaving the sector entirely. Several interviewees had recently entered their job from another part of government where they pioneered analytics, or were about to leave from their current post. One modeller described how their manager called them \myenquote{the single point of failure for the entire force} [\informant{police-w1}]. As discussed above, there is significant concern within the sector that less resourced sister organisations will import the models without the hard-won practices to understand and mitigate issues such as bias and discrimination. Some of the informal practices that are established might be able to be documented, at least for inspiration if not for reproduction---employee handover notes are of course commonplace in these organisations. Yet other practices, particularly any critical skills that led to the establishment of practices in the first place, will likely be more challenging to codify.

Encoding social practices that surround software systems has always been challenging. The stakes are now higher than ever. Relevant efforts might involve the creation of informal and dynamic knowledgebases and virtual communities to share ethical issues and quandaries in relation to algorithmic support in practice~\cite{vealebinns}, but expecting this to arise organically in competitive or resource-scarce fields is risky. Considering what collaboration in these domains could and should look like is of immediate importance to practitioners today.

\subsection{`Looks like we've 100\% accuracy': Talking performance}

Some of the most value laden aspects of machine learned models relate to loss functions and performance metrics. Yet, beyond accuracy, false positives or negatives, it fast becomes difficult to explain performance effectively to those lacking technical background, but whose vertical accountability for the project or necessary, extensive domain knowledge makes it necessary. As recalled above, some informants complained of challenges explaining performance when accuracy was not the appropriate task-specific metric, such as in heavily imbalanced datasets (where you can get a high accuracy by using a dumb classifier that always predicts one class). There are a range of performance metrics suitable for imbalanced data~\cite{japkowicz2011evaluating}, but these mostly lack clear analogies for laypeople. Moving away from binary classification, explaining performance metrics for continuous regression tasks or multiple classification tasks is arguably more challenging still.  

In other cases described above, performance was judged in other ways: models were not trusted or thought valuable if they did not contain individuals' \myenquote{preferred risk indicators}  [\informant{taxlead2}]; were too similar to analysis that existed before [\informant{police-b3}]; or even if they were \emph{more} accurate than was initially planned for, as the commissioners would rather the rest of that performance be substituted for interpretability [\informant{police-c1}]. Other informants emphasised the importance of talking to users before determining performance metrics [\informant{police-b4}], as in some cases only actionable knowledge is worth optimising for (see also~\cite{adepeju2016}). This broadly chimed with many respondents' conception of the most important performance metric of all---for contractors, whether a client bought a model, and for public servants or in-house modellers, whether their department actually used it.

Given that performance metrics are one of the most value-laden parts of the machine learning process~\cite{barocas2016big,coglianese2016regulating}, it will be key to discuss them both with statistical rigour and with practical relevance. This intuitively seems to present domain-specific challenges in training, visualisation, user interfaces, statistics and metrics, problem structuring and knowledge elicitation, among other fields.

\section{Concluding remarks}

Researchers should be wary of assuming, as seems often the case in current discourse, that those involved in the procurement and deployment of these systems are necessarily na\"{i}ve about challenges such as fairness and accountability in the public sector's use of algorithmic decision support. This assumption sits particularly uncomfortably with the value attributed to participatory design and action research in HCI and information systems~\cite{hayes2011relationship,baskerville1996critical}. While those involved in acquiring these technologies for the public sector might not be prime candidates for developing new statistical technologies for understanding bias and outputs in complex models, this does not mean that they do not care or do not try to tackle ethical issues that they perceive. Indeed, as well as the individual perspectives in this paper, some public agencies are already developing their own in-house ethical codes for data science activities~\cite{GDSethics2015}. Yet issues like fairness have been shown to come with technically difficult to reconcile, or even irreconcilable trade-offs---something well-demonstrated by Alexandra Chouldechova's impossibility theorem illustrating that independently plausible formal definitions of fairness can be statistically incompatible with one another~\cite{Chouldechova_2017}, or concerns raised that explanation facilities might work better for some outputs than for others~\cite{edwardsveale}. Reconciling these harder boundaries and issues within messy organisational contexts will present a major challenge to research uptake in this field in the coming years.

Where to go from here? We believe that the challenges we outlined above---dealing with changing data, better understanding discretion and the augmentation of model outputs, better transmission of social practices and improved communication of nuanced aspects of performance---sit amongst a range of promising areas for future interdisciplinary collaboration. The implicit and explicit assumptions of proposed solutions to both these challenges and to the broader issues must be stress-tested in real situations. This presents important questions of methodology. Domain-specific, organisational and contextual factors are crucial to closely consider in the context of interventions intended to improve the fairness and accountability of algorithmic decision-support. The institutional constraints, high stakes and crossed lines of accountability in the public sector arguably presents even more reason to do so. Only so much can be learned from studying systems \emph{in vitro}, even with access to impressive quantities of relevant, quality data with which to experiment. Those interested in transformative impact in the area of fair and accountable machine learning must move towards studying these processes \emph{in vivo}, in the messy, socio-technical contexts in which they inevitably exist. Interventions will have to cope with institutional factors, political winds, technical lock-in and ancient, withering infrastructure head on, as they would have to in the real world. Researchers will have to facilitate the navigation of contested values, and will not always have the freedom of seeking the types of accountability or fairness that they feel most comfortable with. Such challenges should be embraced. To enable this, trust will need to be built between public bodies and researchers; trust that is currently being endangered by `gotcha!'--style research that seeks to identify problematic aspects of algorithmic systems from afar without working collaboratively to understand the processes by which they came about and might be practically remedied. Action research is a core methodology that would support these aims~\cite{baskerville1996critical}, but the combination of high stakes and a wariness that researchers might be spending more effort looking for algorithmic harms than offering help to fix it might make public agencies reluctant to open up to research interventions.

Rarely have the issues HCI concerns itself with been as directly involved in steering choices related to the use of governmental power as much as they are today. As we involve more advanced decision-support, and even decision-making, systems in the workings of the state, this field might even be the `difference that makes a difference' to the rights and freedoms of vulnerable societal groups. We believe that making this difference is possible, but only in close collaboration with different disciplines, practitioners and affected stakeholders. Future research must engage with not only with the new questions and avenues of exploration such research brings, but also the practical constraints that come with studying politically charged settings and developing workable social and technical improvements within them.

\section{Acknowledgments}
Funded by the Engineering \& Physical Sciences Research Council (EPSRC): EP/M507970/1 [MV], EP/J017728/2 [MVK, RB]. This study was approved by UCL's Research Ethics Committee (7617/001) and is distributed CC-BY 3.0. Participants in several workshops---FAT/ML '17 (Halifax, CA); Politiques de mod\'elisation algorithmique (ULB, BE); TRILCon (Winchester, UK); Big Data: New Challenges for Law \& Ethics (Ljubljana, SI); and The Human Use of Machine Learning (Ca'Foscari, Venice, IT)---provided helpful comments on presentations of early forms of this work.

% BALANCE COLUMNS
\balance{}

% REFERENCES FORMAT
% References must be the same font size as other body text.
\bibliographystyle{SIGCHI-Reference-Format}
\bibliography{chi-bibliography}

\end{document}